\documentclass[a4paper,12pt]{article}

\pdfoutput=1

\usepackage{amsmath}
\usepackage{amssymb}
\usepackage{mathrsfs}
\usepackage{dsfont}
\usepackage{bbm}
\usepackage{graphicx,subfigure}
\usepackage[numbers,sort&compress]{natbib}
\usepackage[nodisplayskipstretch]{setspace}
\renewcommand{\arraystretch}{0.75}

\usepackage{xcolor}
\usepackage{ulem}
\usepackage{multirow}

\numberwithin{equation}{section}

\newlength{\dinwidth}
\newlength{\dinmargin}
\newcommand{\abs}[1]{\left\lvert #1 \right\rvert}

\def\fb{\,{\rm fb}}
\def\eV{\,{\rm eV}}
\def\MeV{\,{\rm MeV}}
\def\GeV{\,{\rm GeV}}
\def\TeV{\,{\rm TeV}}
\def\SM{{\rm SM}}
\def\Zp{{Z^\prime}}

\newcommand{\del}[1]{\textcolor{gray}{\sout{#1}}}

\makeatletter
\newcommand{\thickhline}{%
    \noalign {\ifnum 0=`}\fi \hrule height 1pt
    \futurelet \reserved@a \@xhline
}
\makeatother

\begin{document}

\title{\bf Constraints on $\boldsymbol{\Zp}$ Boson \\
within Minimal Flavor Violation}

\author{
C.S. Kim\footnote{cskim@yonsei.ac.kr},
~Xing-Bo Yuan\footnote{xbyuan@yonsei.ac.kr}~
and
Ya-Juan Zheng\footnote{yjzheng@yonsei.ac.kr}
}

\date{}
\maketitle

\vspace{-3.6em}

\begin{center}
\textit{Department of Physics and IPAP, Yonsei University, Seoul 120-749, Korea}\\[0.5em]
\end{center}

\vspace{1em}

\begin{abstract}                
{\noindent}We explore a $\Zp$ boson coupled only with the Standard Model (SM) fermions ($\Zp f\bar{f}$) in the framework of Minimal Flavor Violation. We study its effects on the processes with lepton flavor violation $\ell_j\to \ell_i\ell_k\bar{\ell}_l$, $\ell_j\to\ell_i\gamma$, $\mu^{-}N\to e^- N$, quark flavor changing neutral currents $b\to s \ell^+\ell^-$, neutral $B$ and $K$ meson mixing, and $e^+e^-\to f\bar{f}$ at the LEP experiment to constrain the parameter space of $\Zp$ mass and couplings. We find that among those relevant processes, $\mu\to e\gamma$, $\mu\to 3e$, $\mu$-$e$ conversion and $e^+e^-\to f\bar{f}$ can put more stringent bounds on $\Zp f\bar{f}$ couplings normalized by $\Zp$ mass. Its implications on various processes are obtained, such as $B$ and $K$ mixing and $B/K\to M \ell_1 \bar\ell_2$ decays. In addition, we also make analysis on $\Zp$ signatures at the LHC with $\sqrt{s}=8$ and $13\TeV$.
%We find that In this framework, the $\Zp$ is favored in the mass region $>$ ?? TeV.
\end{abstract}

\newpage

\section{Introduction}

An additional spin-1 neutral gauge boson called $\Zp$ is known to appear in many scenarios beyond the Standard Model (BSM)~\cite{Langacker:2008yv}, such as grand unified models~\cite{Hewett:1988xc}, superstring-inspired models~\cite{Komachenko:1989qn}, models with extra dimensions~\cite{Randall:1999ee}, \textit{etc}.
Discovery of a $\Zp$ boson at present and future high energy colliders could be one of the best illuminating signatures of BSM physics~\cite{Nath:2010zj}. Experimental searches for a massive spin-1 resonance have been performed at the LEP~\cite{Schael:2013ita}, Tevatron~\cite{Abe:1997fd} and the current LHC~\cite{Aad:2014cka,Khachatryan:2014fba} experiments.

Interactions of $\Zp$ to the SM fermions depend on the parameters of specific models. An interesting case is that its interactions with the SM fermions are family nonuniversal~\cite{Chaudhuri:1994cd}. A general theoretical framework for a family nonuniversal $\Zp$ boson has been investigated in Ref.~\cite{Langacker:2000ju}, and the flavor changing effects in such scenario in both quark~\cite{Langacker:2000ju, Barger:2003hg} and lepton sectors~\cite{Chiang:2011cv,Crivellin:2015era,Becirevic:2016zri,Yue:2016mqm} have been extensively explored. However, as is well known, the induced quark flavor-changing neutral currents (FCNC) mediated by $\Zp$ boson are phenomenologically dangerous.
Potentially large contributions to FCNC in quark sector appear in many BSM scenarios, \textit{e.g.}, two Higgs doublet model, technicolor, \textit{etc.}, which may result in severe phenomenological difficulties~\cite{Browder:2008em}.
To avoid such large FCNC effects, solutions~\cite{Glashow:1976nt,Branco:1996bq,Ko:2012hd} such as the Natural Flavor Conservation hypothesis~\cite{Glashow:1976nt} and the BGL model~\cite{Branco:1996bq} have been  invented.
In this work we want to investigate the FCNC effects within the hypothesis of Minimal Flavor Violation (MFV)~\cite{Chivukula:1987py,Hall:1990ac,D'Ambrosio:2002ex}. In the MFV hypothesis, it is assumed that all flavor violating currents at low energy are controlled by the Yukawa couplings, so that all the FCNC interactions in quark sector are naturally suppressed by the Cabibbo-Kobayashi-Maskawa (CKM) \cite{CKM} factors, as in the SM. It is also possible to extend the MFV hypothesis to leptonic sector~\cite{Cirigliano:2005ck, Davidson:2006bd, Gavela:2009cd}.
 %,Alonso:2011jd,Sierra:2012yy,Branco:2006hz
 However, the leptonic MFV is quite ambiguous because of the unknown mechanism responsible for the origin of neutrino mass. For a recent review on the MFV in both quark and lepton sectors, we refer to Ref.~\cite{Isidori:2012ts}. The MFV hypothesis can be used in effective field theory to perform model-independent studies on possible BSM effects~\cite{D'Ambrosio:2002ex}. One can also implement MFV in a renormalizable model, which result in the SM-like flavor and CP violation at low energies~\cite{Dimou:2015yng}.

%In ref.~\cite{Dimou:2015yng}, MFV is realized in the GUT model with a $S_4\times U(1)$ symmetry. see~\cite{Dimou:2015yng}'s introduction. and reference therein.

In this paper, we investigate such a possibility that a $\Zp$ boson couples to quark and lepton with flavor violating interactions within the most general MFV hypothesis. Instead of building a specific model, we adopt a bottom-up approach, where all the $\Zp$ interactions arise from some effective operators which involve both the SM fermions and a $\Zp$ boson, and at the same time satisfy the criterion of the MFV hypothesis. In quark sector, this scenario has been studied in the case of the SM $Z$ boson with modified couplings to down-type quarks~\cite{Guadagnoli:2013mru}. In our work, the MFV hypothesis is extended to both quark and lepton sector with the implementation of $\Zp$ boson. Its effects on various processes such as lepton flavor violating (LFV) transitions $\ell_j\to \ell_i\ell_k\bar{\ell}_l$, $\ell_j\to\ell_i\gamma$ and $\mu^{-}N\to e^- N$, quark FCNC processes in neutral $B$ and $K$ meson mixing, and high-energy collisions $e^+ e^- \to f \bar f$ at LEP are investigated in detail. Constraints on the $\Zp$ mass and its couplings to fermions are derived. Its implications to the LHC direct searches at $\sqrt{s}=8\TeV$ (LHC run I) and $13\TeV$ (LHC run II) are also discussed.

It is also noted that current LHCb run I data present some deviations from the SM predictions~\cite{Langenbruch:2015dqz}. The measured ratio $R_K\equiv \mathcal B(B \to K \mu^+ \mu^-)/ \mathcal B ( B \to K e^+ e^-)$ shows $2.6\,\sigma$ departure from unity~\cite{Aaij:2014ora}, which may indicate violation of lepton universality. In addition, some angular observables (mainly so-called $P_5^\prime$) in $B \to K^* \mu^+ \mu^-$ decay differ from the SM predictions with a significance of about $3\,\sigma$~\cite{Aaij:2013qta}.
Many BSM scenarios are proposed to explain such anomalies~\cite{Gauld:2013qba,Altmannshofer:2014cfa,Gauld:2013qja,Glashow:2014iga}, most of which contain a $\Zp$ boson.
A general feature presented in these SM extensions is the $\Zp$ couplings with charged leptons and down-type quarks are typically family nonuniversal to explain the observed $R_K$ and the anomalies in $B \to K^* \mu^+ \mu^-$ decay, respectively. Similar features also present in the couplings within the MFV hypothesis. Therefore, the MFV $\Zp$ boson could be a candidate to explain the current LHCb anomalies, which will be investigated in detail here.

The paper is organized as follows: In section~\ref{sec:MFV}, we briefly describe the MFV hypothesis and introduce theoretical framework for an MFV $\Zp$ boson. Some relevant processes in both quark and lepton sector are discussed in section~\ref{sec:theo}. In section~\ref{sec:analysis}, numerical analysis is performed. Then, we give predictions on both low and high energy processes in section~\ref{sec:prediction} and conclude in section~\ref{sec:conclusion}.

\section{Minimal Flavor Violating $\boldsymbol{\Zp}$ Boson}\label{sec:MFV}

In the absence of Yukawa interactions, the SM quark sector exhibits a global flavor symmetry~\cite{Chivukula:1987py}
\begin{align}
  G_{\rm QF} \equiv SU(3)_{Q_L} \otimes SU(3)_{U_R}\otimes SU(3)_{D_R},
\end{align}
plus three additional $U(1)$ groups identified as baryon number, hypercharge and the Peccei-Quinn symmetry~\cite{Peccei:1977hh}. The SM quark sector contains one $SU(2)_L$ doublet $Q_L$ and two $SU(2)_L$ singlets $U_R$ and $D_R$, all of which consist of three families. Under the flavor symmetry $G_{\rm QF}$, they transform as
\begin{align}
  Q_L \sim {\bf (3,1,1)}, \qquad U_R \sim {\bf (1,3,1)}, \qquad D_R \sim {\bf (1,1,3)}.
\end{align}
The MFV hypothesis assumes that the dynamics of flavor and CP violation at low energy is determined by the structure of the Yukawa couplings~\cite{D'Ambrosio:2002ex}. Technically, the flavor symmetry group $G_{\rm QF}$, which is explicitly broken by the Yukawa couplings $Y_U$ and $Y_D$ in the SM, can be formally recovered by promoting the Yukawa couplings to be spurion fields with the transformation property~\cite{D'Ambrosio:2002ex}
\begin{align}
  Y_U \sim {\bf (3,\bar 3,1)},\qquad {\rm and}\qquad Y_D \sim {\bf (3,1,\bar 3)}.
\end{align}
Then it is possible to construct $G_{\rm QF}$ invariant effective operators from the SM fields and the spurions $Y_U$ and $Y_D$, which could satisfy the criterion of MFV.

The MFV hypothesis can also be extended to the lepton sector. However, the mechanism responsible for neutrino masses are unknown at present. Thus, there is no unique way to introduce the MFV principle in the lepton sector. Various definitions of lepton MFV have been proposed in the literature~\cite{Cirigliano:2005ck, Davidson:2006bd,Gavela:2009cd}, which depend on the specific BSM scenarios generating the sources of lepton flavor symmetry breaking, such as seesaw mechanism~\cite{seesaw1,seesaw12,seesaw2,seesaw3}. Here, we consider the realization of leptonic MFV within the so-called minimal field content~\cite{Cirigliano:2005ck} with one left-handed lepton doublet $L_L$ and one right-handed singlet $e_R$. The lepton flavor symmetry is~\cite{Chivukula:1987py}
\begin{align}
  G_{\rm LF}\equiv SU(3)_L \otimes SU(3)_E
\end{align}
plus two $U(1)$ symmetries respecting lepton number $U(1)_{\rm LN}$ and the weak hypercharge. The Yukawa interaction, which generates lepton mass and breaks the lepton flavor symmetry, reads~\cite{Cirigliano:2005ck}
\begin{align}
  \Delta\mathcal L =&-\bar e_R \lambda_e H^\dagger L_L -\frac{1}{2\Lambda_{\rm LN}} \left(\bar L_L^c \tau_2 H\right)g_\nu \left(H^T \tau_2 L_L\right) +\rm h.c. \nonumber\\
\xrightarrow{\rm sym.br.}& -v\bar e_R \lambda_e e_L -\frac{v^2}{2\Lambda_{\rm LN}} \bar \nu_L^c g_\nu \nu_L +\rm h.c.,
\end{align}
where $\Lambda_{\rm LN}$ denotes the scale of the lepton number symmetry $U(1)_{\rm LN}$  breaking, and the vacuum expectation value $v=174\GeV$. The charged lepton and neutrino Yukawa couplings $\lambda_e$ and $g_\nu$ are $3 \times 3$ matrices in flavor space. In this case, the tiny neutrino masses are explained by the smallness of $v/\Lambda_{\rm LN}$.

Considering effective couplings of a $\Zp$ boson to the SM fermions, the relevant effective operators satisfying the MFV hypothesis can be written as
\begin{align}\label{eq:operator}
  \mathcal O_L^q=\left(\bar Q_L \Delta_q \gamma^\mu Q_L \right) Z_\mu^\prime\qquad {\rm and} \qquad
  \mathcal O_L^\ell=\left(\bar L_L \Delta_\ell \gamma^\mu L_L \right) Z_\mu^\prime .
\end{align}
In order to make them invariant under the quark and lepton flavor group $G_{\rm QF}$ and $G_{\rm LF}$, the coupling matrices should have the form as
\begin{align}\label{eq:coupling}
  \Delta_q= \kappa_0 \mathds{1} + \kappa_1 Y_U Y_U^\dagger + \dotsc,
  \qquad
  {\rm and}
  \qquad
  \Delta_\ell= \lambda_0 \mathds{1} + \lambda_1 g_\nu^\dagger g_\nu+\dotsc,
\end{align}
where $\mathds 1$ denotes $3\times 3$ identity matrix in flavor space. In the series, $\kappa_i$ and $\lambda_i$ are unknown real coefficients, and the terms with higher orders of the spurions $Y_U$, $Y_D$, $\lambda_e$ and $g_\nu$ are indicated by the ellipses. As in Ref.~\cite{Faller:2013gca}, the flavor conserving term $\mathds 1$ is also considered.

In the literature, there are some other treatments which can be used to realize MFV hypothesis for the $\Zp$ couplings. In the expansion series Eq.~(\ref{eq:coupling}), higher order terms can be resumed by the Cayley-Hamilton identity and the series stop at the order $\kappa_2 (Y_UY_U^\dagger)^2$ and $\lambda_2 (g_\nu^\dagger g_\nu)^2$ after neglecting the down-type fermion Yukawa couplings~\cite{Colangelo:2008qp}. For quark sector, it is also possible to use the approach of nonlinear parameterization to account for the higher order contributions~\cite{Feldmann:2008ja}. In addition, the operators with right-handed fields $u_R$, $d_R$ and $e_R$ can also be constructed to satisfy MFV hypothesis. However, the corresponding flavor violating couplings are suppressed by small down-type fermion Yukawa couplings such as $\lambda_b$ and $\lambda_\mu$. In this work, we concentrate on a $\Zp$ boson in which its interactions with fermions satisfy the MFV hypothesis and originate from the effective operators $\mathcal O_L^q$ and $\mathcal O_L^\ell$ in Eq.~(\ref{eq:operator}). Our analysis can be straightforwardly extended to more general cases, such as including some of the above ingredients or considering  lepton MFV with seesaw mechanism.

In the following analysis, it is convenient to work in the Lagrangian,
\begin{align}\label{eq:Lagrangian:Zp}
  \mathcal L=\Gamma_{\ell\ell^\prime}^L \left( \bar \ell \gamma^\mu P_L \ell^\prime \right) Z_\mu^\prime+\Gamma_{q q^\prime}^L \left ( \bar q \gamma^\mu P_L q^\prime \right) Z_\mu^\prime + \left( L\rightarrow R \right),
\end{align}
where $P_{L,R}=(1\mp \gamma_5)/2$, $q$ and $q^\prime$ ($\ell$ and $\ell^\prime$) denote a pair of up- or down-type quarks (leptons). Then, the MFV operators of Eq.~(\ref{eq:operator}) result in the following couplings
\begin{align}
  \Gamma_{\ell\ell^\prime}^L&=\lambda_0 \delta_{\ell\ell^\prime} +\lambda_1 \frac{\Lambda_{\rm LN}^2}{v^4}\sum_{\nu_i} m_{\nu_i}^2 U_{\ell \nu_i}U_{\ell^\prime \nu_i}^*,& \Gamma_{\ell\ell^\prime}^R&=0,
  \nonumber\\
  \Gamma_{qq^\prime}^L&= \kappa_0 \delta_{qq^\prime} + \kappa_1 \lambda_t^2 V_{tq}^* V_{tq^\prime} ,& \Gamma_{qq^\prime}^R&=0,
\end{align}
where $m_\nu$ denotes diagonal neutrino mass matrix, and $\hat U$ the Pontecorvo-Maki-Nakagawa-Sakata (PMNS) matrix~\cite{Pontecorvo:1967fh}. Numerically, the diagonal elements of the lepton coupling matrix $\Gamma_{\ell \ell^\prime}^L$ are almost universal. Therefore, we define a new coupling $\bar\lambda=\Gamma_{ee}^L$ and take the approximation $\Gamma_{\ell\ell}^L\approx \bar\lambda$ in the following discussion. In the case of normal hierarchy (NH) of neutrino mass spectrum, assuming $m_1=0.2\eV$ and $\Lambda_{\rm LN}=10^{14}\GeV$, the coupling matices read
\begin{align}
  \label{eq:Gamma:num}
  \abs{\Gamma_{\ell \ell^\prime}^L}&= \abs{\bar\lambda} \mathds{1}+\abs{\lambda_1}
  \begin{pmatrix}
    0 & 0.23 & 0.31\\
    0.26 & 0 & 1.28\\
    0.31 & 1.28 & 0
  \end{pmatrix}
  \times 10^{-2},
  \nonumber\\[0.5em]
  \abs{\Gamma_{qq\prime}^L}&= \abs{\kappa_0} \mathds{1} + \abs{\kappa_1}
  \begin{pmatrix}
    0.00007 & 0.00031 & 0.000760 \\
    0.00031 & 0.00144 & 0.003475 \\
    0.00760 & 0.03575 & 0.886868
  \end{pmatrix}.
\end{align}
The flavor conserving couplings are almost universal in lepton sector but rather hierarchical in quark sector. For flavor changing couplings, they are suppressed in both quark and lepton sectors, which are the feature of MFV hypothesis.

As in many BSM scenarios, the mass of $\Zp$ boson is a free parameter in our case. In this work, we focus on a TeV scale $\Zp$ boson, which may explain some current observed anomalies and could be detected at the LHC.

\section{Processes to Constrain the Parameter Space of $\boldsymbol{\Zp}$ }\label{sec:theo}

Due to its family nonuniversal couplings, an MFV $\Zp$ boson may affect processes from low-energy flavor transitions all the way to high-energy collider processes. The most relevant processes of leptonic decays $\ell_j \to \ell_i \ell_k \bar \ell_l$, $\ell_j\to \ell_i \gamma$, $\mu$-$e$ conversion in the lepton sector, quark FCNC processes $b\to s\ell^+\ell^-$, neutral $B$ and $K$ meson mixing $B_s$-$\bar B_s$, $B_d$-$\bar B_d$ and $K^0$-$\bar K^0$ processes in the quark sector, and $e^+e^-\to f\bar{f}$ at the LEP experiment are investigated in detail in this section.

\subsection{Lepton flavor violation processes}\label{sec:LFV}

\subsubsection{leptonic decays $\boldsymbol{\ell_j \to \ell_i \ell_k \bar \ell_l}$}

Among LFV decays, the most important processes contain the decay of a charged lepton $\ell_j$ into three charged leptons $\ell_i$, $\ell_k$ and $\bar \ell_l$, \textit{e.g.}, $\mu\to 3e$. With the Lagrangian~\eqref{eq:Lagrangian:Zp}, the tree-level $\Zp$ exchange results in the following branching ratios~\cite{Langacker:2000ju,Chiang:2011cv,Crivellin:2015era}
\begin{align}
  \mathcal B(\ell_j \to \ell_i \ell_k \bar \ell_l) &= \frac{\tau_jm_j}{1536\pi^3}\left(\frac{m_j}{m_\Zp}\right)^4 \left(
      \abs{\Gamma_{ij}^L\Gamma_{kl}^L + \Gamma_{kj}^L\Gamma_{il}^L}^2 + \abs{\Gamma_{ij}^L\Gamma_{kl}^R}^2 +\abs{\Gamma_{kj}^L\Gamma_{il}^R}^2 + \bigl(L\leftrightarrow R\bigr)
      \right),
\nonumber\\
  \mathcal B(\ell_j \to \ell_i \ell_i \bar \ell_l) &= \frac{\tau_jm_j}{1536\pi^3}\left(\frac{m_j}{m_\Zp}\right)^4 \left(
     2 \abs{\Gamma_{ij}^L\Gamma_{il}^L}^2 +   \abs{\Gamma_{ij}^L\Gamma_{il}^R}^2
     + \bigl(L \leftrightarrow R\bigr)
     \right),
\end{align}
which are applied to the case that the two same sign final leptons with same and different flavor, respectively.

\subsubsection{leptonic decays $\boldsymbol{\ell_j\to \ell_i \gamma}$}

Another relevant decay is loop induced radiative decay $\ell_j \to \ell_i \gamma$, \textit{e.g.}, $\mu \to e \gamma$. After neglecting the mass of final leptons, the branching ratio reads~\cite{Chiang:2011cv,Crivellin:2015era}
\begin{align}
  \mathcal B(\ell_j \to \ell_i \gamma)  = \frac{\alpha_e\tau_jm_j}{9(4\pi)^4}\left(\frac{m_j}{m_\Zp}\right)^4
 \biggl (\biggl\lvert\sum_k \Gamma_{jk}^L \Gamma_{ki}^L - \frac{3 m_k}{m_j}\Gamma_{kj}^L\Gamma_{ki}^R \biggr\rvert^2 + \bigl(L\leftrightarrow R \bigr) \biggr),
\end{align}
where the enhancement factor $m_k/m_j$ is similar to its counterpart in quark sector $b \to s \gamma$ decay.

\subsubsection{$\boldsymbol{\mu}$-$\boldsymbol{e}$ conversion}

For the $\mu$-$e$-$\Zp$ coupling in particular, a strong bound comes from $\mu$-$e$ conversion in nuclei. The experimental sensitivities are expected to be improved by several orders of magnitude and reach about $\mathcal O (10^{-17})$ in near future~\cite{Uchida:2014jda}. The branching fraction for $\mu$-$e$ conversion in atomic nuclei $N$ reads~\cite{Kuno:1999jp}
\begin{align}
  \mathcal B(\mu^- N \to e^- N)= \frac{\alpha_e^3  m_\mu^5}{(8\pi)^2 \Gamma_{\rm capt}}\frac{\abs{F_p}^2}{m_\Zp^4}&\bigl( \abs{\Gamma_{e\mu}^L}^2 +\abs{\Gamma_{e\mu}^R}^2 \bigr) \\
  &\times \bigl\lvert (2Z+N)(\Gamma_{uu}^L + \Gamma_{uu}^R) + (Z+2N)(\Gamma_{dd}^L + \Gamma_{dd}^R) \bigr\rvert^2\nonumber,
\end{align}
where $Z$ and $N$ denote the atomic and neutron number respectively. $\Gamma_{\rm capt}$ denotes the $\mu$ capture rate, $Z_{\rm eff}$ the effective atomic number, and $F_p$ the nuclear matrix element~\cite{Kitano:2002mt}. Unlike other LFV processes, $\mu$-$e$ conversion in nuclei involves interactions with light quarks, which could constrain the flavor conserving $u$-$u$-$\Zp$ and $d$-$d$-$\Zp$ type couplings.

\subsection{Quark flavor changing neutral current processes}

\subsubsection{$\boldsymbol{\abs{\Delta F}=1}$ transition: $\boldsymbol{b\to s \ell^+ \ell^-}$ processes}\label{sec:btosll}

Generally, the effective Hamilton for $b \to s \ell^+ \ell^-$ transitions can be written as~\cite{Bobeth:1999mk}
\begin{align}
  \mathcal H_{\rm eff}^{\Delta F=1}=-\frac{4G_F}{\sqrt 2} V_{tb}V_{ts}^* \sum_{i=1}^{10} \mathcal C_i \mathcal O_i + h.c.,
\end{align}
plus small $\mathcal O (V_{ub}V_{ts}^*)$ corrections, where explicit expressions of the four-quark operators $\mathcal O_{1-6}$ can be found in Ref.~\cite{Bobeth:1999mk}. In the SM, the electromagnetic dipole operator and semileptonic four-fermion operators play a leading role~\cite{Ali:1999mm}
\begin{align*}
  \mathcal O_{7\gamma} = \frac{e}{16\pi^2}m_b \bigl(\bar s \sigma_{\mu\nu} P_R b \bigr) F^{\mu\nu},
  \quad
  \mathcal O_{9\ell} = \frac{e^2}{16\pi^2}  \bigl(\bar s \gamma_\mu P_L b\bigr) \bigl(\bar \ell \gamma^\mu \ell\bigr),
  \quad
  \mathcal O_{10\ell} = \frac{e^2}{16\pi^2} \bigl(\bar s \gamma_\mu P_L b \bigr) \bigl(\bar \ell \gamma^\mu \gamma_5 \ell\bigr).
\end{align*}
In the Lagrangian Eq.~\eqref{eq:Lagrangian:Zp}, the Wilson coefficient $\mathcal C_7$ is affected at loop level while the semileptonic operators receive tree-level contributions from $\Zp$ exchange, which result in
\begin{align}
  \begin{pmatrix}
    \mathcal C_{9\ell}^{\rm NP} \\
    \mathcal C_{10\ell}^{\rm NP}
  \end{pmatrix}
=-\frac{\pi}{\sqrt 2 \alpha_e G_F V_{tb}V_{ts}^*}\frac{\Gamma_{sb}^L}{m_\Zp^2}
\begin{pmatrix}
  \Gamma_{\ell\ell}^R+\Gamma_{\ell\ell}^L\\
  \Gamma_{\ell\ell}^R-\Gamma_{\ell\ell}^L
\end{pmatrix}.
\end{align}
After neglecting the right-handed currents, there are only two model-independent parameters $\bigl(\mathcal C_{9e}^{\rm NP}=-\mathcal C_{10e}^{\rm NP}, \mathcal C_{9\mu}^{\rm NP}=-\mathcal C_{10\mu}^{\rm NP} \bigr)$ in $b \to s e^+ e^-$ and $b \to s \mu^+ \mu^-$ transitions, which have been fit to current experimental data by several groups~\cite{Altmannshofer:2014rta, Descotes-Genon:2015uva,Descotes-Genon:2013wba}.

\subsubsection{$\boldsymbol{\abs{\Delta F}=2}$ transition: neutral $B$ and $K$ meson mixing}\label{sec:mixing}

The FCNC processes $B_s$-$\bar B_s$, $B_d$-$\bar B_d$ and $K^0$-$\bar K^0$ mixing play an important role in constraining possible BSM effects. In the SM, $B_s$-$\bar B_s $ mixing occurs via box diagrams by exchanging $W^\pm$ boson. The mixing strength is described by the mass difference $\Delta m_{s} = 2|\langle B_s | \mathcal H^{\Delta B=2} | \bar B_s \rangle | $ governed by the effective Hamiltonian~\cite{Buras:2001ra}
\begin{align}
\mathcal H_{\rm eff}^{\Delta B=2} =\frac{G_F^2}{16\pi^2}m_W^2 (V_{tb}^{}V_{ts}^*)^2  \mathcal C_1^{\rm VLL} \bigl(\bar s^\alpha \gamma_\mu P_L b^\alpha\bigr) \bigl (\bar s^\beta \gamma^\mu P_L b^\beta \bigr )   +h.c..
\end{align}
The Wilson coefficient $\mathcal C^{\rm VLL}_1$ at matching scale $\mu=\mu_W$ can be found in Ref.~\cite{Buchalla:1995vs}  and its QCD renormalization group evolution to $B$ meson scale can be found in Refs.~\cite{Buras:2001ra,Ciuchini:1997bw}. In our scenario, the left-handed current can modify the Wilson coefficient at high scale as~\cite{Buras:2012fs}
\begin{align}\label{eq:WC:mixing}
\mathcal C_{1,\,\rm NP}^{\rm VLL}=\frac{16\pi^2}{G_F}\frac{1}{m_W^2 (V_{tb}^{}V_{ts}^*)^2} \frac{(\Gamma_{sb}^L)^2}{2m_\Zp^2}.
\end{align}
Similar expressions hold for $B_d$-$\bar B_d$ and $K^0$-$\bar K^0$ mixing in both SM and BSM physics. In particular, the $K^0$-$\bar K^0$ mixing is more complicated. From the effective Hamiltonian, one can build two observables, mass difference $\Delta m_K$ and CP-violating parameter $\varepsilon_K$~\cite{Buras:2013ooa}. However, compared to $B$ meson mixing, these two observables suffer from large theoretical uncertainties, especially for $\Delta m_K$~\cite{Buras:2013ooa,Bertolini:2014sua}. The uncertainties from short-distance and long-distance contributions to the mass difference have been discussed in Refs.~\cite{Herrlich:1996vf,Herrlich:1993yv,Brod:2011ty,Brod:2010mj} and Refs.~\cite{Antonelli:1996qd,Bertolini:1997ir,Buras:2010pza,Buras:2014maa}, respectively. Recent lattice QCD calculations can be found in Refs.~\cite{Christ:2012se,Blum:2014tka}. We refer to Ref.~\cite{Amhis:2015xea} for a recent review on $B$ and $K$ meson mixing.

It is also noted that, the Wilson coefficients of Eq.~\eqref{eq:WC:mixing} should run to the low scales $\mu_K=2\GeV$ for $K$ mixing and $\mu_B$ for $B$ mixing under QCD renormalization group evolution. The particular low scale value should match the evaluation scale of the corresponding hadronic matrix element. The evolution from high scale to low scale should be done with the changing of the effective flavors $n_f=6 \to 4$ for $K$ mixing and $n_f=6 \to 5$ for $B$ mixing. All the relevant formulae can be found in Refs.~\cite{Buras:2000if,Buras:2001ra,Buras:2012fs,Ciuchini:1997bw}.

\subsection{$e^+ e^- \to f \bar f$ at the LEP}

The LEP-II $e^+ e^- \to f \bar f$ data, where $f$ denotes a quark or lepton flavor, were taken at the energies $\sqrt s$ increasing from $130\GeV$ to $209\GeV$~\cite{Schael:2013ita}. The cross sections and forward-backward asymmetries for various fermion pairs can be used to search for a TeV scale $\Zp$ boson. As a model-independent approach, the LEP collaboration uses the following effective Lagrangian, \textit{i.e.}, contact interaction to constrain possible BSM effects~\cite{Eichten:1983hw}
\begin{align}\label{eq:Lagrangian:Lambda}
  \mathcal L_{\rm eff}=\frac{4\pi}{(1+\delta_f)\Lambda_{f,\,\pm}^2}\sum_{i,j=L,R} \eta_{ij}\bigl(\bar e_i \gamma_\mu e_i\bigr) \bigl(\bar f_j \gamma^\mu f_j \bigr),
\end{align}
where $\delta_f=1$ $(0)$ for $f=e$ $(f\neq e)$. The free parameters $\Lambda_{f,\,\pm}$ encode possible BSM effects, which may constructively ($+$) or destructively ($-$) interfere with the SM contributions. Interactions with different chiralities and interferences correspond to the choices of $\eta_{ij}=\pm 1$ or $0$.  With the notation in Eq.~\eqref{eq:Lagrangian:Zp}, the scale $\Lambda_{f,\,\pm}$ and parameters $\lambda_{ij}$ read
\begin{align}\label{eq:Lambda}
  \Lambda_{f,\,\pm}=\left(\frac{4\pi m_\Zp^2}{\bigl\lvert \Gamma_{ee}^L\Gamma_{ff}^L \bigr\rvert} \right)^{1/2} \qquad {\rm and} \qquad
  \eta_{ij}=\Biggl\lbrace
\begin{aligned}
   -& {\rm sgn}\bigl(\Gamma_{ee}^L\Gamma_{ff}^L\bigr),  \quad& &i=j=L,\\
     & 0, \quad&& {\rm others}.
\end{aligned}
\end{align}

In the case of hadron final states, since it is difficult to distinguish final jets originated from different flavors, the LEP collaboration interprets the experimental data in several cases. We adopt the interpretation that possible new interactions only exist between electrons and a single up-type flavor. Since only $u$ and $c$ quarks can be produced at the LEP energies and the $\Zp$ couplings to them are almost universal, the LEP lower bound $\Lambda_{uu,\,\pm}^{\rm LEP}$~\cite{Schael:2013ita} can be converted to $\Lambda_{u,\,\pm}>\Lambda_{uu,\,\pm}^{\rm LEP}/\sqrt 2$ without loss of generality.

In the case of $f=\ell=e,\mu,\tau$, additional $u$- and $t$- channel diagrams with LFV couplings also contribute to $\Lambda_{\ell,\,\pm}$. However, these LFV couplings are highly suppressed as shown in Eq.~\eqref{eq:Gamma:num}. Thus, Eq.~\eqref{eq:Lambda} for $\Lambda_{\ell,\,\pm}$ holds in a good approximation.

\subsection{Other relevant processes}

In this part, we discuss briefly about some other processes receiving contributions from the MFV $\Zp$ boson.
Due to its couplings to muons and neutrinos, the $\Zp$ boson can contribute to neutrino trident production  $\nu_\mu N \to \nu N \mu^+ \mu^-$~\cite{Altmannshofer:2014cfa,Crivellin:2015era,Glashow:2014iga}.
Using combined measurements from CHARM-II~\cite{Geiregat:1990gz}, CCFR~\cite{Mishra:1991bv} and NuTeV~\cite{Adams:1999mn}, a bound on $\bar\lambda$ and $\lambda_1$ is derived, which turned out to be much weaker than the one from the LFV decays $\ell_j \to \ell_i \ell_k \bar\ell_l$ and $\ell_j \to \ell_i \gamma$.
Similarly, except for the $\tau \to \mu \nu \bar \nu$ decay, the leptonic decays $\ell_j\to \ell_i \nu \bar\nu$
\footnote{For $\mu \to e \nu \bar \nu$ decay, to avoid large corrections to the Fermi constant $G_F$, we demand $\abs{\mathcal B(\mu \to e \nu \bar \nu)_{\rm exp}-\mathcal B(\mu\to e \nu \bar\nu)_{\SM}}<4\times 10^{-5}$ as suggested in Ref.~\cite{Crivellin:2015era}.}
can not put further constraints on the model parameters.
For $\tau \to \mu \nu \bar \nu$ in particular, its experimentally measured branching ratio is currently more than $2\,\sigma$ above the SM prediction~\cite{Altmannshofer:2014cfa}. The allowed parameter space from this process is not compatible for the ones from other $\ell_j \to \ell_i \nu \bar\nu$ modes. As described in section.~\ref{sec:MFV}, the right-handed $\Zp$ couplings is not included in this scenario. Therefore, the MFV $\Zp$ considered in this paper can not contribute to leptonic electric dipole moments~\cite{Chiang:2011cv}. Furthermore, the $\Zp$ effects on the anomalous magnetic moment $a_\mu$ is always negative~\cite{Crivellin:2015era} and can not relax the longstanding discrepancy between the SM and experiment~\cite{Jegerlehner:2009ry}. At last,  the bounds from conversion $\mu^- e^+ \to \mu^+ e^-$~\cite{Willmann:1998gd} are also very weak.

\section{Numerical Analysis and Discussions}\label{sec:analysis}

\begin{table}[t]
\renewcommand{\arraystretch}{1.2}
  \centering
  \begin{tabular}{l | l l l}
    \thickhline
    \multirow{6}{*}{\rotatebox[origin=c]{90}{lepton sector}}& $\sin^2\theta_{12}$ & $0.308_{-0.017}^{+0.017}$& \cite{Capozzi:2013csa}
    \\
    & $\sin^2\theta_{23}$ & $0.437_{-0.023}^{+0.033}$  $(0.455_{-0.031}^{+0.039})$ & \cite{Capozzi:2013csa}
    \\
    & $\sin^2\theta_{13}$ & $0.0234_{-0.0019}^{+0.0020}$ $(0.0240_{-0.0022}^{+0.0019})$ &\cite{Capozzi:2013csa}
    \\
    & $\delta/\pi$ & $1.39_{-0.27}^{+0.38}$ $(1.31_{-0.33}^{+0.29})$& \cite{Capozzi:2013csa}
    \\
    &  \,$\Delta m_{21}^2$ $[10^{-5} \,{\rm eV}^2]$ & $7.54_{-0.22}^{+0.26}$ & \cite{Capozzi:2013csa}
    \\
    & $\abs{\Delta m^2}$ $[10^{-3}\,{\rm eV}^2]$ & $2.43_{-0.06}^{+0.06}$ $(2.38_{-0.06}^{+0.06})$ & \cite{Capozzi:2013csa}
    \\\hline
    \multirow{7}{*}{\rotatebox[origin=c]{90}{quark sector}}&  $|V_{us}| f_+^{K\to \pi}(0)$  & $0.21664\pm 0.00048 $ & \cite{Charles:2015gya}
    \\
    &  $|V_{ub}|$ (semi-leptonic) & $(3.70\pm 0.12 \pm 0.26) \times 10^{-3}$ & \cite{Charles:2015gya}
    \\
    &  $|V_{cb}|$ (semi-leptonic) & $(41.0 \pm 0.33 \pm 0.74) \times 10^{-3}$ & \cite{Charles:2015gya}
    \\
    & $\gamma$ $ [^\circ]$ & $73.2_{-7.0}^{+6.3}$ & \cite{Charles:2015gya}
    \\
    & $\overline{m}_c (\overline{m}_c)$ & $(1.286 \pm 0.013 \pm 0.040) \, {\rm GeV}$ & \cite{Charles:2015gya}
    \\
    & $\overline{m}_b (\overline{m}_b)$ & $(4.18 \pm 0.03 ) \, {\rm GeV}$ & \cite{Agashe:2014kda}
    \\
    &  $\overline{m}_t (\overline{m}_t)$ & $(165.95 \pm 0.35 \pm 0.64) \, {\rm GeV}$ & \cite{Charles:2015gya}
    \\\hline
    \multirow{12}{*}{\rotatebox[origin=c]{90}{$B$ and $K$ meson mixing}} &  $f_+^{K\to \pi}(0)$ & $0.9641 \pm 0.0015 \pm 0.0045$& \cite{Charles:2015gya}
    \\
    &  $f_K$ & $(155.2 \pm 0.2 \pm 0.6)\MeV$ & \cite{Charles:2015gya}
    \\
    &  $f_{B_s}$               & $(225.6 \pm 1.1 \pm 5.4)\MeV$ & \cite{Charles:2015gya}
    \\
    &  $f_{B_s}/f_{B_d}$     & $1.205 \pm 0.004 \pm 0.007$   &\cite{Charles:2015gya}
    \\
    &  $\hat{B}_K$  & $0.7615\pm 0.0027 \pm 0.0137$ & \cite{Charles:2015gya}
    \\
    &  $\hat{B}_{B_s}$       & $1.320\pm 0.017 \pm 0.030$ & \cite{Charles:2015gya}
    \\
    &  $\hat{B}_{B_s}/\hat{B}_{B_d}$ & $1.023 \pm 0.013 \pm 0.014$ & \cite{Charles:2015gya}
    \\
    &  $\eta_{cc}$ & $1.87 \pm 0.76$ & \cite{Brod:2011ty}
    \\
    &  $\eta_{ct}$ & $0.497 \pm 0.047$ & \cite{Brod:2010mj}
    \\
    &  $\eta_{tt}$ & $0.5765 \pm 0.0065$ & \cite{Herrlich:1993yv}
    \\
    &  $\kappa_\epsilon$ & $0.940 \pm 0.013 \pm 0.023$ & \cite{Buras:2010pza}
    \\
    &  $\varphi_\epsilon$ & $(43.51 \pm 0.05)^\circ$ & \cite{Buras:2010pza}
    \\\thickhline
  \end{tabular}
  \caption{\baselineskip 3.0ex
    Input parameters used in the numerical analysis for the lepton sector, quark sector and $B$ and $K$ meson mixing. The mixing parameters in the lepton sector(values in brackets) correspond to NH (inverted hierarchy (IH)).}
  \label{tab:input}
\end{table}

\begin{table}[t]
\renewcommand{\arraystretch}{1.2}
  \centering
  \begin{tabular}{l l l l l}
    \thickhline
    OBSERVABLE & SM & EXP & \\
    \hline
    ${\cal B}(\mu\to e\gamma)\vphantom{\frac{1}{2}_|^|}$ &~~-& $<5.7\times10^{-13}$ &\cite{Agashe:2014kda}
    \\
    ${\cal B}(\mu{\rm Ti}\to e{\rm Ti})\vphantom{\frac{1}{2}_|^|}$ &~~-& $<6.1\times10^{-13}$& \cite{Papoulias:2013gha}~
    \\
    ${\cal B}(\mu{\rm Au}\to e{\rm Au})\vphantom{\frac{1}{2}_|^|}$ &~~-& $<7.0\times10^{-13}$ &\cite{Agashe:2014kda}
    \\
    ${\cal B}(\mu^-\to e^-e^-e^+)\vphantom{\frac{1}{2}_|^|}$&~~-& $<1.0\times10^{-12}$ &\cite{Agashe:2014kda}
    \\ \hline
    ${\cal B}(\tau\to\mu\gamma)\vphantom{\frac{1}{2}_|^|}$ &~~-& $<4.4\times10^{-8}$ &\cite{Agashe:2014kda}
    \\
    ${\cal B}(\tau^-\to\mu^-\mu^-\mu^+)\vphantom{\frac{1}{2}_|^|}$ &~~-& $<2.1\times10^{-8}$ &\cite{Agashe:2014kda}
    \\
    ${\cal B}(\tau^-\to\mu^-e^-e^+)\vphantom{\frac{1}{2}_|^|}$ &~~-& $<1.8\times10^{-8}$ &\cite{Agashe:2014kda}
    \\ \hline
    ${\cal B}(\tau\to e\gamma)\vphantom{\frac{1}{2}_|^|}$ &~~-& $<3.3\times10^{-8}$ &\cite{Agashe:2014kda}
    \\
    ${\cal B}(\tau^-\to e^-e^-e^+)\vphantom{\frac{1}{2}_|^|}$&~~-&  $<2.7\times10^{-8}$& \cite{Agashe:2014kda}
    \\
    ${\cal B}(\tau^-\to e^-\mu^-\mu^+)\vphantom{\frac{1}{2}_|^|}$ &~~-& $<2.7\times10^{-8}$ &\cite{Agashe:2014kda}
    \\\hline
    $\mathcal B (K_L \to e^\pm \mu^\mp)$ &~~-& $<4.7 \times 10^{-12}$ & \cite{Agashe:2014kda}
    \\\hline
    $\Delta m_d [\,{\rm ps}^{-1}]$ & ~\,$0.51 \pm 0.06 $& ~\,$0.510 \pm 0.003 $ & \cite{Amhis:2014hma}
    \\
    $\Delta m_s [\,{\rm ps}^{-1}]$ &  $16.93 \pm 1.16$ & $17.757 \pm 0.021 $ & \cite{Amhis:2014hma}
    \\
    $\Delta m_K [\,10^{-3}\,{\rm ps}^{-1}]$ & ~\,$4.40\pm 1.77$ &~\,$5.293 \pm 0.009$& \cite{Agashe:2014kda}
    \\
    $\abs{\varepsilon_K} [10^{-3}]$ & ~\,$2.10\pm 0.30$ & ~\,$2.228\pm 0.011$& \cite{Agashe:2014kda}
    \\\thickhline
  \end{tabular}
  \caption{\baselineskip 3.0ex
  The SM predictions and experimental measurements for the observables used in the numerical analysis. The upper limits for LFV decays are values corresponding to 90\% CL. }
  \label{tab:exp}
\end{table}

With the theoretical framework described in previous sections, we proceed to present our numerical analysis and discussions in this section. Table~\ref{tab:input} shows the input parameters for various processes mentioned above. In Table~\ref{tab:exp}, we summarize the SM predictions and current experimental data for these processes. The theoretical uncertainties of the observables in $B$ and $K$ meson mixings are obtained with varying each input parameter within $1\,\sigma$ range and adding each individual uncertainty in quadrature.

As discussed in section~\ref{sec:MFV}, the relevant model parameters in our case contain the flavor conserving (changing) couplings $\kappa_0$ ($\kappa_1$) and $\bar\lambda$ ($\lambda_1$), which correspond to quark and lepton sector respectively, and the $\Zp$ mass $m_\Zp^{}$. Since we concentrate on a TeV scale $\Zp$ boson, the mass effects are decoupled for all the processes except those at the LHC. Therefore, we choose the model parameter as
\begin{align}
  \left(\frac{\kappa_0}{m_\Zp},\frac{ \kappa_1}{m_\Zp},\frac{ \bar\lambda}{m_\Zp}, \frac{\lambda_1}{m_\Zp}\right).
\end{align}
The constraints on these parameters will be discussed in the following sections.

Compared to the lepton processes, the quark FCNC processes still suffer from large theoretical uncertainty due to hadronic inputs. In order to derive allowed parameter space from these processes, we impose the experimental constraints in the same way as in Ref.~\cite{Jung:2012vu}, \textit{i.e.}, for each point in the parameter space, a theoretical range is constructed from the prediction of the observable in that point together with the corresponding theoretical uncertainty. If this range overlaps with the $2\,\sigma$ range of the experimental measurement, then this point is regarded as allowed. To be conservative, the theoretical uncertainty is taken as twice the one listed in Table~\ref{tab:exp}. Since the main theoretical uncertainties arise from hadronic input parameters, which are common to both the SM and the MFV $\Zp$ boson, the relative theoretical uncertainty is assumed to be constant over the whole parameter space.

\subsection{Bounds on $\Zp$ couplings to leptons}

\begin{figure}[t]
  \centering
  \subfigure{\includegraphics[width=0.45\textwidth]{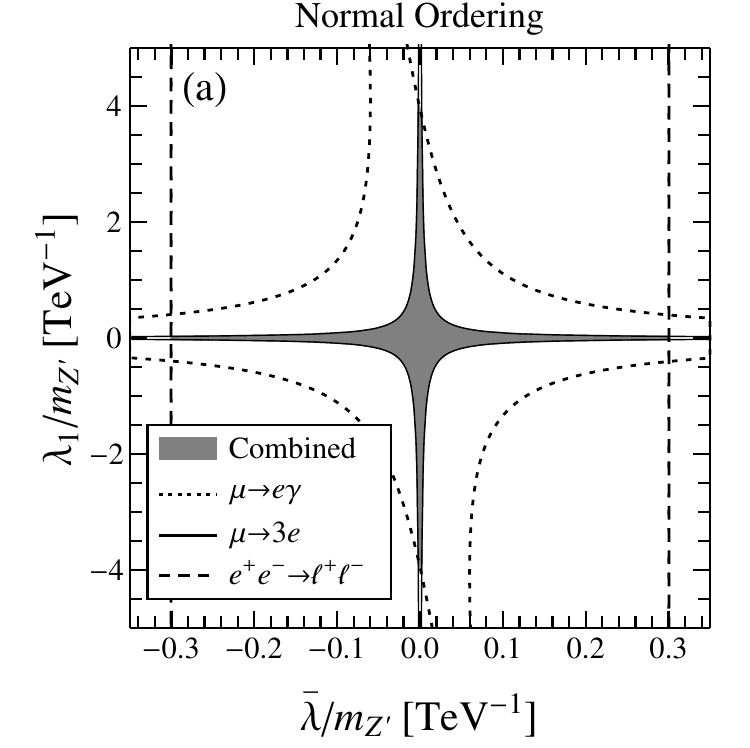}}
  \subfigure{\includegraphics[width=0.45\textwidth]{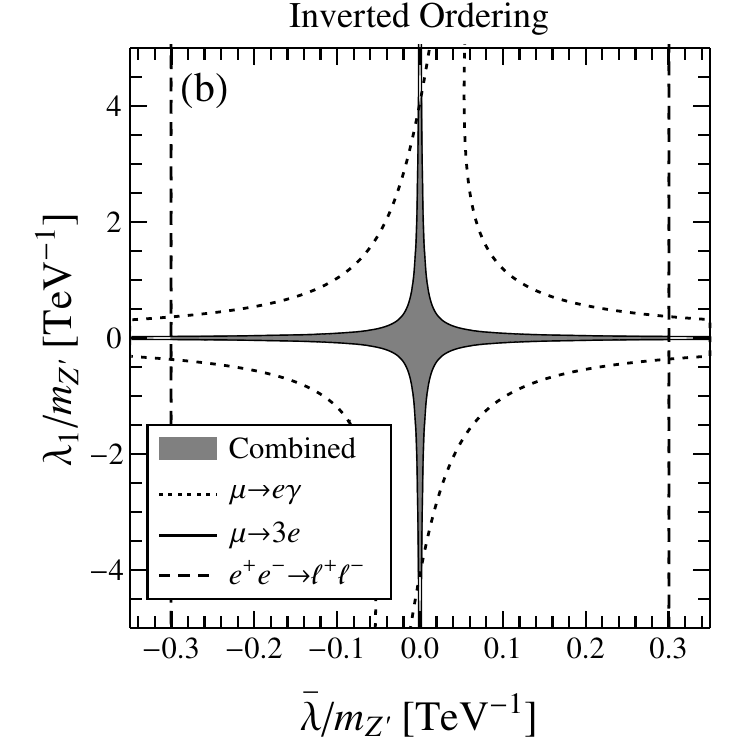}}
  \caption{\baselineskip 3.0ex
    Combined constraints on the $\Zp$ parameters, plotted in $(\bar\lambda/m_\Zp, \lambda_1/m_\Zp)$ plane. (a) and (b) denotes the results with NH and IH of neutrino masses, respectively. The gray region indicates the allowed parameter space after combing all the relevant processes, which are dominated by $\mu \to e \gamma$ (dotted), $\mu \to 3e$ (solid), and $e^+e^- \to \ell^+ \ell^-$ at LEP (dashed).}
\label{fig:PS:lam}
\end{figure}

The processes of lepton radiative decays $\ell_j \to \ell_i \ell_k \bar\ell_l$ (\textit{e.g.}, $\mu \to 3 e$) and $\ell_j \to \ell_i \gamma$ (\textit{e.g.}, $\mu \to e \gamma$), collider processes $e^+ e^- \to \ell^+ \ell^-$ at LEP, neutrino trident production $\nu_\mu N \to \nu N \mu^+ \mu^-$, and $\mu^- e^+ \to \mu^+ e^-$ conversion involve only the $\Zp$ couplings to leptons $\bar\lambda$ and $\lambda_1$, which controls flavor conserving and flavor changing current respectively. After considering the current experimental data of these processes, which are listed in Table~\ref{tab:exp}, it is found that the bounds on the $\Zp$ couplings are dominated by $\mu \to e \gamma$, $\mu \to 3e$ and the LEP processes $e^+ e^- \to \ell^+ \ell^-$. For NH and IH cases, Fig.~\ref{fig:PS:lam} shows the allowed parameter space in $(\bar\lambda / m_\Zp,\lambda_1 / m_\Zp)$ plane.
We can see that the upper bound on $\bar \lambda$ and $\lambda_1$ is provided by $e^+ e^- \to \ell^+ \ell^-$ and a combination of $\mu \to 3e $ and $\mu \to e \gamma$ decays, respectively.
 The latter process also puts a bound on the product of these two couplings, numerically as $\bar\lambda\lambda_1/m_\Zp^2 \lesssim 0.01 \TeV^{-2}$ for NH. Since the bounds from NH and IH are quite similar,  we will only consider the NH case in the following analysis for simplicity.

% $\ell_j \to \ell_i \nu \bar \nu$ (e.g. $\tau \to \mu \nu \bar\nu$)

\subsection{Bounds on $\Zp$ couplings to quarks}

Constraints on the $\Zp$ couplings with quarks come from $B_s$-$\bar B_s$, $B_d$-$\bar B_d$ and $K^0$-$\bar K^0$ mixing. Due to relatively large theoretical uncertainties in $K^0$-$\bar K^0$ mixing discussed in section~\ref{sec:mixing}, we adopt the conservative treatment in Ref.~\cite{Bertolini:2014sua}, \textit{i.e.}, NP contributions to $\Delta m_K$ is available within 50\% range of $\Delta m_K^{\rm exp}$ and $\abs{\varepsilon_K}$ is allowed to vary within a 20\% symmetric range. The experimental measurements on the observables in $B$ and $K$ mixing are listed in Table~\ref{tab:exp}, which show good agreement with the SM predictions. Therefore, a stringent bound on the $\Zp$ parameter is found
\begin{align}
  \abs{\kappa_1/m_\Zp}<0.18\TeV^{-1},
\end{align}
which is dominated by $\Delta m_s$ and slightly stronger than those from $\abs{\varepsilon_K}$ and $\Delta m_d$. Our numerics agree with the fit on the scale $\Lambda$ of $\Delta F=2$ MFV effective operators~\cite{Bona:2007vi}.
%\tgray{[B meson leptonic decays]}

\subsection{Bounds on $\Zp$ couplings with both lepton and quark}

\begin{figure}[t]
  \centering
  \subfigure{\label{fig:PS_kap0_lamb}\includegraphics[width=0.45\textwidth]{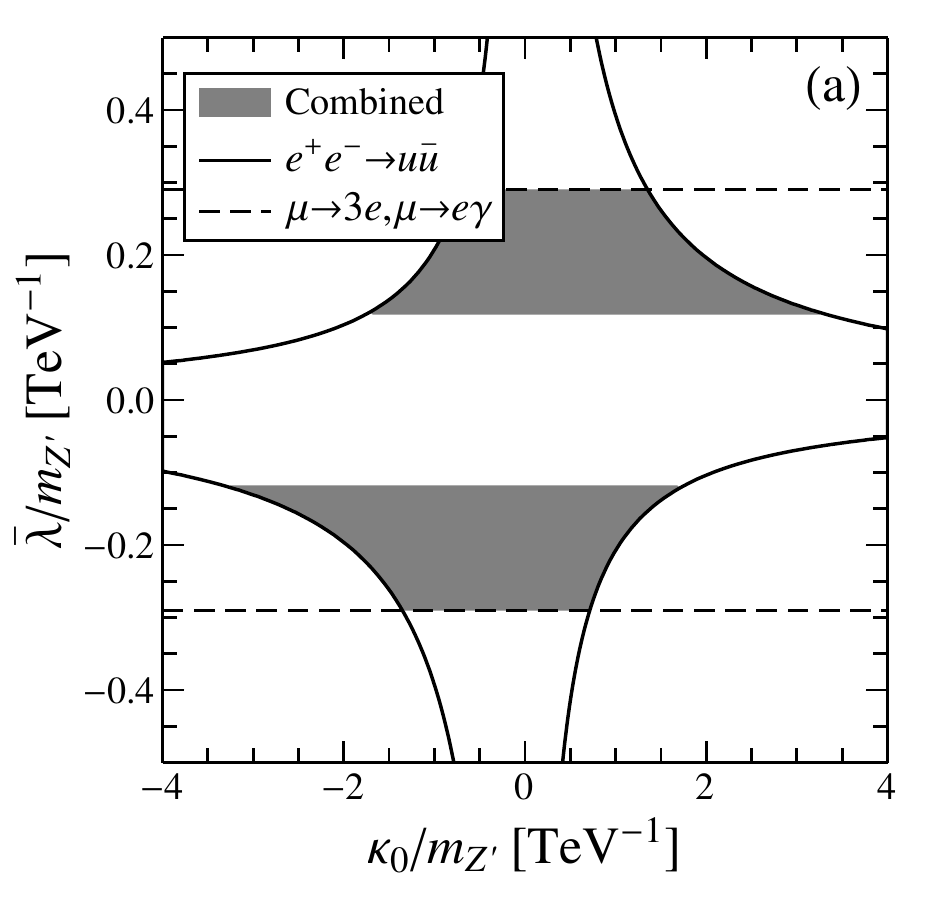}}
  \subfigure{\label{fig:PS_kap1_lamb}\includegraphics[width=0.45\textwidth]{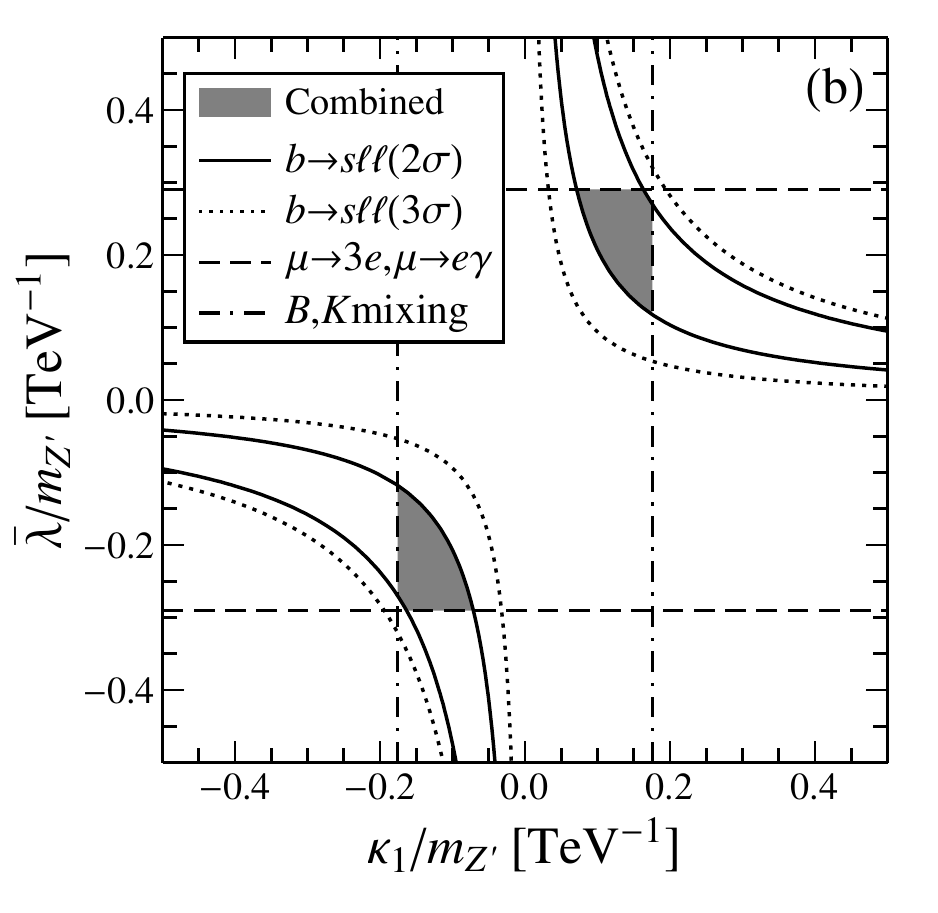}}
  \\
  \subfigure{\label{fig:PS_kap0_lam1}\includegraphics[width=0.45\textwidth]{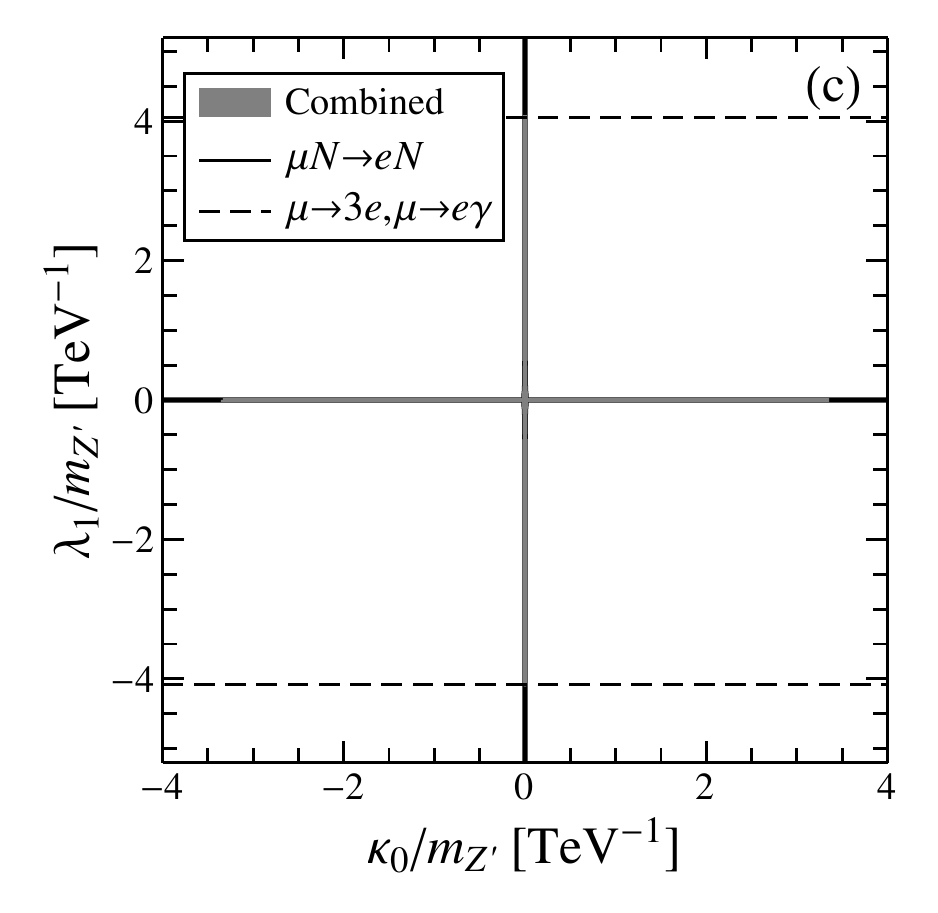}}
  \subfigure{\label{fig:PS_kap1_lam1}\includegraphics[width=0.45\textwidth]{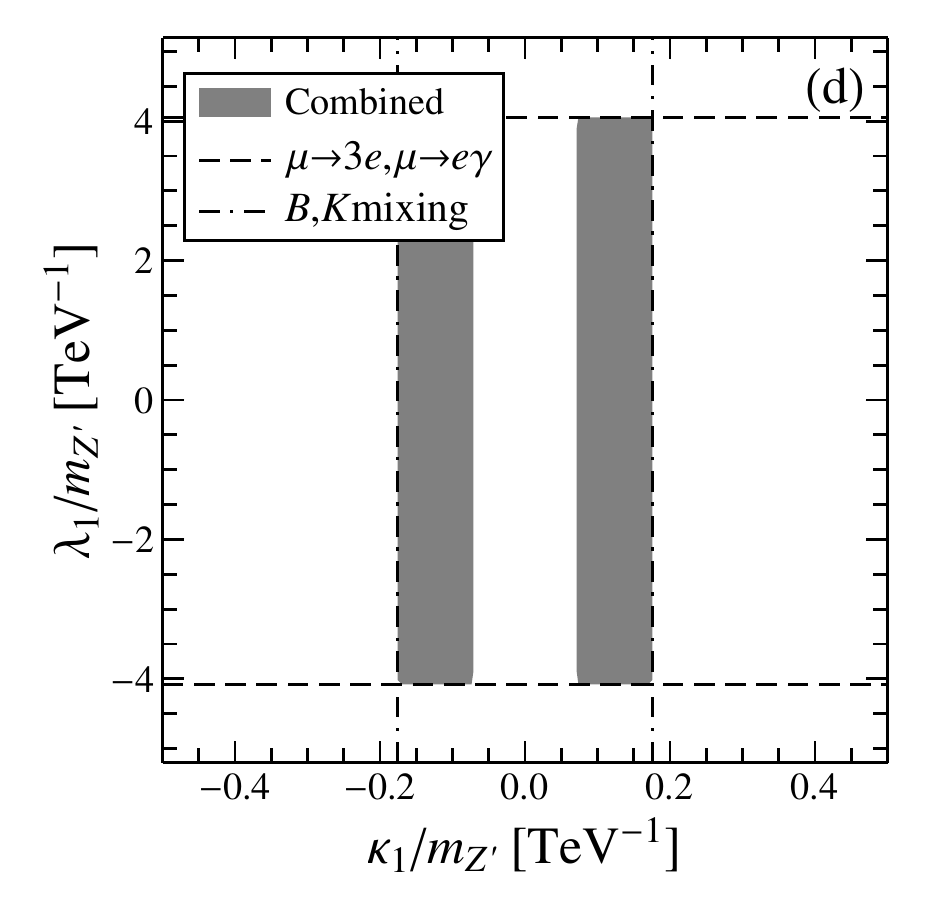}}
  \caption{\baselineskip 3.0ex
  Constraints for the $\Zp$ couplings in the planes of paris of $\kappa_0,\kappa_1,\bar\lambda$ and $\lambda_1$. Combined allowed regions are shown in gray. The various lines indicate individual bounds described in the legend.
  }
  \label{fig:PS_kap_lam}
\end{figure}

 From section~\ref{sec:btosll}, $b \to s \ell^+ \ell^-$ processes involve both $\Zp$ couplings to quarks and leptons, which appear as $\kappa_1\bar\lambda$ in the amptulide. In the experimental side, recent measurements on these processes have shown deviations from the SM~\cite{Aaij:2015esa,Langenbruch:2015dqz}. For example, angular observable $P_5^\prime$~\cite{Matias:2012xw} in $B \to K^* \mu^+ \mu^-$ decay exhibits large deviations from the SM predictions in some bins~\cite{Aaij:2013qta}. The ratio $R_K \equiv \mathcal B (B \to K \mu^+ \mu^-)/\mathcal B(B \to K e^+ e^-)$ measured by LHCb shows $2.6\,\sigma$ discrepancy from unity, which is predicted by the SM with very good accuracy, and may give a hint of lepton flavor non-universality~\cite{Aaij:2014ora}. After model-independent global fits, BSM interpretations have been investigated by several groups~\cite{Altmannshofer:2014rta,Descotes-Genon:2015uva,Descotes-Genon:2013wba}. In the following analysis, we adopt the recent results in Ref.~\cite{Descotes-Genon:2015uva}, which include all available $b\to s \mu^+ \mu^-$~\cite{Aaij:2013qta,Aaij:2014pli,Aaij:2015esa} and $b \to s e^+ e^-$~\cite{Aaij:2013hha,Aaij:2014ora,Aaij:2015dea} data. With the two model-independent Wilson coefficients $\bigl(\mathcal C_{9e}^{\rm NP}=-\mathcal C_{10e}^{\rm NP}, \mathcal C_{9\mu}^{\rm NP}=-\mathcal C_{10\mu}^{\rm NP} \bigr)$, the current anomalies can be explained with a non-vanishing contribution to the muon sector $\mathcal C_{9\mu}^{\rm NP}\approx -1$ but non-significant NP contribution in the electron sector. In this case, the significance for deviation from lepton flavor universality is about $1.2\,\sigma$.

As discussed in section~\ref{sec:LFV}, the collider processes $e^+ e^- \to q \bar q$ at LEP and $\mu^- N \to e^- N$ conversion can constrain the productions of $\Zp$ couplings $\kappa_0\lambda_1$ and $\kappa_0\bar\lambda$, respectively. After considering the bounds obtained in the last two subsections, combined allowed regions as well as bounds from individual processes are shown in Fig.~\ref{fig:PS_kap_lam}. From these plots, we observe that
\begin{itemize}
\item As shown in $(\kappa_1,\bar\lambda)$ plane of Fig.~\ref{fig:PS_kap1_lamb}, the current $b \to s \ell^+ \ell^-$ anomalies can be explained by an MFV $\Zp$ boson at $2\,\sigma$ level after considering the constraints from $B$ and $K$ mixing as well as the combined constraints from all the lepton processes. In the solution, both flavor changing coupling $\kappa_1$ and flavor conserving coupling $\bar\lambda$ have a lower bound, which would result in non-vanishing effects on the LFV decays as well as $B$ and $K$ mixing.
\item In $(\kappa_0, \bar\lambda)$ plane of Fig.~\ref{fig:PS_kap0_lamb}, combination of $e^+ e^- \to u \bar u$ and $b \to s \ell^+ \ell^-$ processes put upper bounds on $\kappa_0$, which is not constrained by other individual process. Since the allowed regions in this plane deviate from $\bar\lambda=0$ axis, most parts of the parameter space may suggest signatures for $pp \to \Zp \to \ell^+ \ell^-$ processes at LHC.
\item In $(\kappa_1, \lambda_1)$ plane of Fig.~\ref{fig:PS_kap1_lam1}, the combined constraints are much stronger than the one from $K_L \to e^\pm \mu^\mp$ decay. Although large $\lambda_1$ is still allowed, $\kappa_1$ is stringently constrained by $B$ and $K$ mixing. Therefore, the resulting bound $\kappa_1\lambda_1/m_\Zp^2< 0.7\TeV^{-2}$ makes the branching ratios of relevant $B$ and $K$ LFV decays not very large.
\item In $(\kappa_0, \lambda_1)$ plane of Fig.~\ref{fig:PS_kap0_lam1}, upper bounds on these two parameters are relatively loose. However, the production $\kappa_0\lambda_1$ is strongly constrained as $\kappa_0\lambda_1/m_\Zp^2< 0.001 \TeV^{-2}$ by $\mu^- {\rm Au} \to e^- {\rm Au}$ due to its tiny experimentally allowed rates. Hence, an MFV $\Zp$ boson almost can not produce LFV dilepton signatures at LHC.
\end{itemize}

%LFV and LFC are typical loose and strong ...

\section{Predictions on Low and High Energy Processes}\label{sec:prediction}

Due to the current anomalies in $b \to s \ell^+ \ell^-$ transitions, some $\Zp$ couplings acquire non-vanishing values after the global fit. In this section, we discuss their impacts on both low and high energy processes.

\subsection{Predictions on low energy flavor processes}
\begin{figure}[t]
  \centering
  \subfigure{\includegraphics[width=0.45\textwidth]{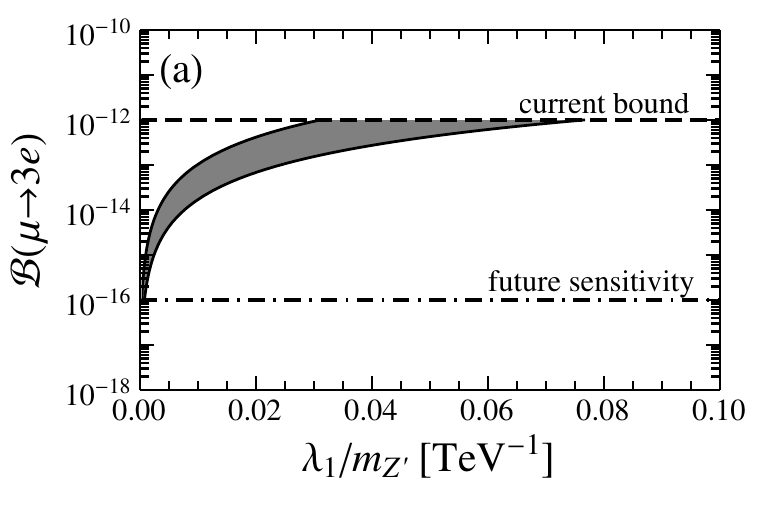}}
  \quad
  \subfigure{\includegraphics[width=0.45\textwidth]{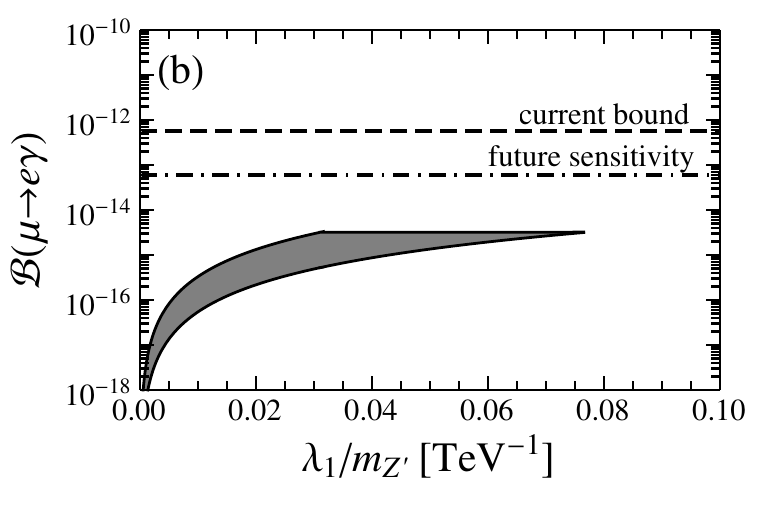}}
  \caption{\baselineskip 3.0ex
    MFV $\Zp$ predictions on $\mathcal B(\mu \to 3e)$ and $\mathcal B ( \mu \to e \gamma)$ as a function of $\lambda_1/m_\Zp$. The allowed range of the branching ratios are shown in gray, which are obtained from the combined allowed regions in Figs.~\ref{fig:PS:lam} and~\ref{fig:PS_kap_lam}. The dashed line denotes the current experimental upper bound, while the dot dashed line is for future sensitivity.}
\label{fig:lepton_LFV_decay}
\end{figure}

As shown in Fig.~\ref{fig:PS_kap1_lamb}, the flavor conserving coupling $\lambda_1$ should be nonzero after considering the $b \to s \ell^+ \ell^-$ processes. Therefore, there exists a lower bound on the branching ratios of $\mu \to 3 e$ and $\mu \to e \gamma$ for a given $\lambda_1$. In the combined allowed parameter space obtained in the previous section, the allowed range of $\mathcal B ( \mu \to 3 e)$ and $\mathcal B (\mu \to e \gamma)$ as a function of $\lambda_1/m_\Zp$ are shown in Fig.~\ref{fig:lepton_LFV_decay}, where future experimental sensitivities on these two decays~\cite{CeiA:2014wea,Baldini:2013ke} are also presented. We note that for $\lambda_1/m_\Zp>0.01 \TeV^{-1}$ the lower limit on the branching ratio is $\mathcal O(10^{-16})$ for $\mu \to e \gamma$ decay, while $\mathcal O(10^{-14})$ for $\mu \to 3 e$ decay, which is about two orders above the future experimental sensitivity. Therefore, $\mu \to 3 e$ decay can be very promising to probe the MFV $\Zp$ effects.

To explain the current $b\to s \ell^+ \ell^-$ anomalies, the FCNC coupling $\kappa_1$ acquire a nonzero value. Since the mixing amplitudes are proportional to $\kappa_1^2$, the $B$ and $K$ mixing may be affected.  In the allowed parameter space of Fig.~\ref{fig:PS_kap1_lam1}, the predictions for various mixing observables are derived
\begin{align}
1.04 <\,  &\Delta m_s / \Delta m_s^\SM   < 1.22, \nonumber\\
1.05 <\, &\Delta m_d / \Delta m_d^\SM < 1.22, \nonumber\\
1.03 <\, & \;\abs{\varepsilon_K} / \abs{\varepsilon_K^\SM} \, < 1.17,
\end{align}
where $\Delta m_K$ is not presented due to minor $\Zp$ effects. All these observables are enhanced by the $\Zp$ effects with about $3\%\sim5$\%. As can be seen in Table~\ref{tab:exp}, the current uncertainties for these mixing observables are dominated by the theoretical calculation, which will be reduced  from the current 10\% to about $3\%\sim 4$\% in the next few years~\cite{Butler:2013kdw}. Therefore, the $B$ and $K$ mixing could have a good opportunity to probe such an MFV $\Zp$ boson.

\begin{figure}[t]
  \centering
  \includegraphics[width=0.9\textwidth]{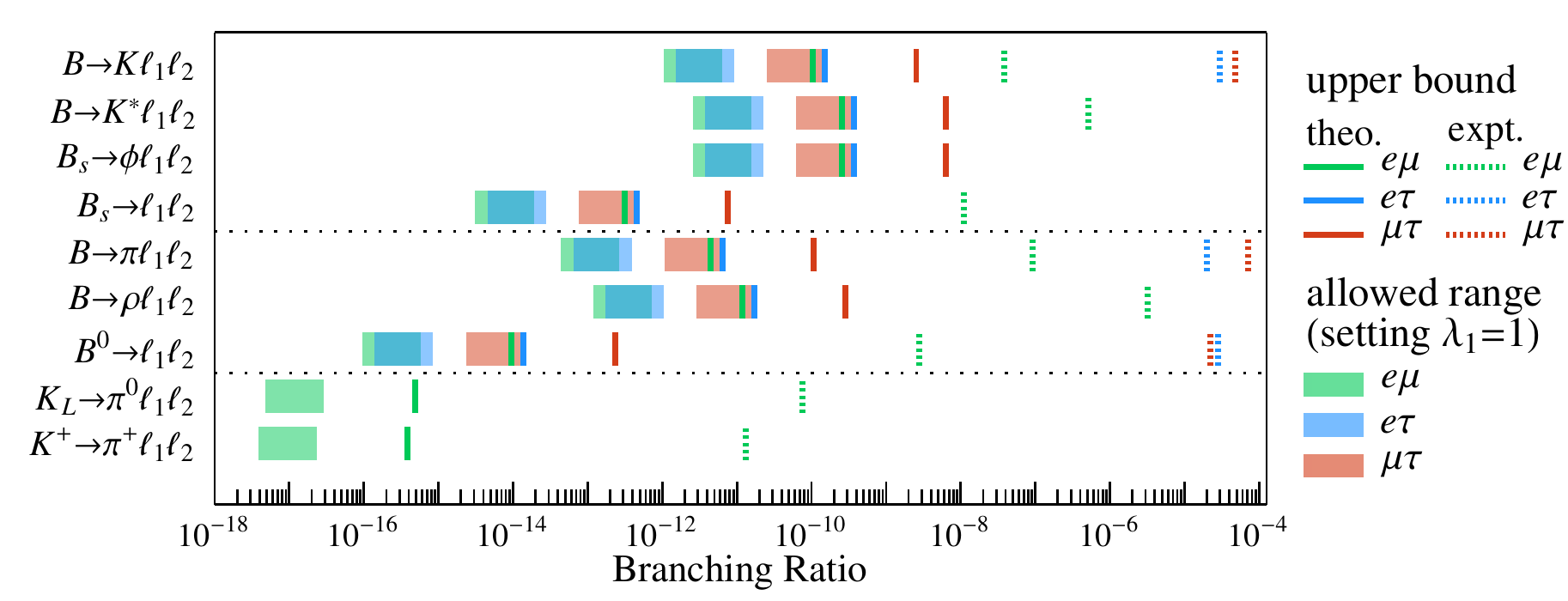}
  \caption{\baselineskip 3.0ex
    Predictions on the branching ratios of various $B$ and $K$ meson LFV decays. The upper bounds from MFV $\Zp$ and current experimental limits are indicated by the solid and dashed line respectively, for the final lepton pairs $e\mu$ (green), $e\tau$ (blue), and $\mu\tau$ (red). In the case of $\lambda_1=1$, the allowed region for the branching ratios are shown in rectangle regions.}
  \label{fig:M_LFV_decay}
\end{figure}

As shown in Fig.~\ref{fig:PS_kap1_lam1}, large values of the LFV coupling $\lambda_1$ is still allowed after the global fit. Together with the non-vanishing coupling $\kappa_1$, they could affect LFV decays of $B$ and $K$ mesons, such as $B \to K^{(*)} e^+ \tau^-$ and $K_L \to \pi^0 e^+ \mu^-$ decays. The branching ratios of these processes are proportional to $\abs{\kappa_1\lambda_1/m_\Zp^2}^2$. Fig.~\ref{fig:M_LFV_decay} shows the predictions on these LFV decays from the allowed $\Zp$ parameter space in Fig.~\ref{fig:PS_kap1_lam1}, as well as the current experimental upper limits~\cite{Agashe:2014kda,Amhis:2014hma}. Since the product $\kappa_1\lambda_1$ is lower bounded for a non-vanishing value of $\lambda_1$, we also give the allowed region of the branching ratios in the case of $\lambda_1/m_\Zp =1\TeV^{-1}$. We can see that the upper limits on the branching ratios are typically  $4 \sim 5$ orders lower than the current observed bounds.

In addition, our predictions are lower than the results from the effective theory analysis with MFV~\cite{Lee:2015qra}. Within an MFV $\Zp$ boson,  both $\kappa_1$ and $\lambda_1$ are bounded from the quark and lepton flavor violating processes, which make the upper limit on the product $\kappa_1\lambda_1$ stronger than that from $K_L \to e^\pm \mu^\mp$ decay. In the effective filed theory approach with MFV, bounds on the effective operators responsible to $B$ and $K$ meson LFV decays mainly arise from $K_L \to e^\pm \mu^\mp$ decay~\cite{Lee:2015qra}. Therefore, its predicted upper bounds on other $B$ and $K$ meson LFV decays are much higher than our results presented in Fig.~\ref{fig:M_LFV_decay}.

\subsection{LHC signatures}

In this part, we first introduce the current measurements and constraints on $\Zp$ boson from $pp$ and $p\bar{p}$ colliders. At the LHC, the CMS collaboration analysed the dimuon (dielectron) mass spectra with the $8\TeV$ run I data corresponding to an integrated luminosity of 20.6 (19.7)$\fb^{-1}$. A Sequential Standard Model (SSM) $Z^{\prime}_{\rm SSM}$ resonance lighter than $2.90\TeV$ is excluded~\cite{Khachatryan:2014fba}. Lower limits on the energy scale parameter for the contact interaction $\Lambda$ are found to be 12.0 (15.2)$\TeV$ for destructive (constructive) interference in the dimuon channel and 13.5 (18.3)$\TeV$ in the dielectron channel~\cite{Khachatryan:2014fba}. The ATLAS collaboration searched for a high-mass resonance decaying into $\tau^+\tau^-$ final state at $\sqrt{s}=8\TeV$ with an integrated luminosity of 19.5-20.3$\fb^{-1}$. Lower mass limit on the $\Zp_{\rm SSM}$ boson is set to be $2.02\TeV$ at $95\%$ C.L.~\cite{Aad:2015osa}. At the Tevatron, both D0 and CDF collaborations searched for a heavy neutral gauge boson in the $e^+e^-$ channel of $p \bar p$ collisions at $ \sqrt{s} = 1.96 \TeV$.  A lower mass limit of about $1\TeV$ for the SSM $\Zp$ boson is presented respectively~\cite{Abazov:2010ti}.  However, the above constraints on the $\Zp$ mass are model dependent, which are typically sensitive to the free parameters such as its couplings to leptons, therefore can still be loosened to some extent.
\begin{figure}[t]
  \centering
  \qquad
  \subfigure{\label{fig:LHC:8}\includegraphics[width=0.45\textwidth]{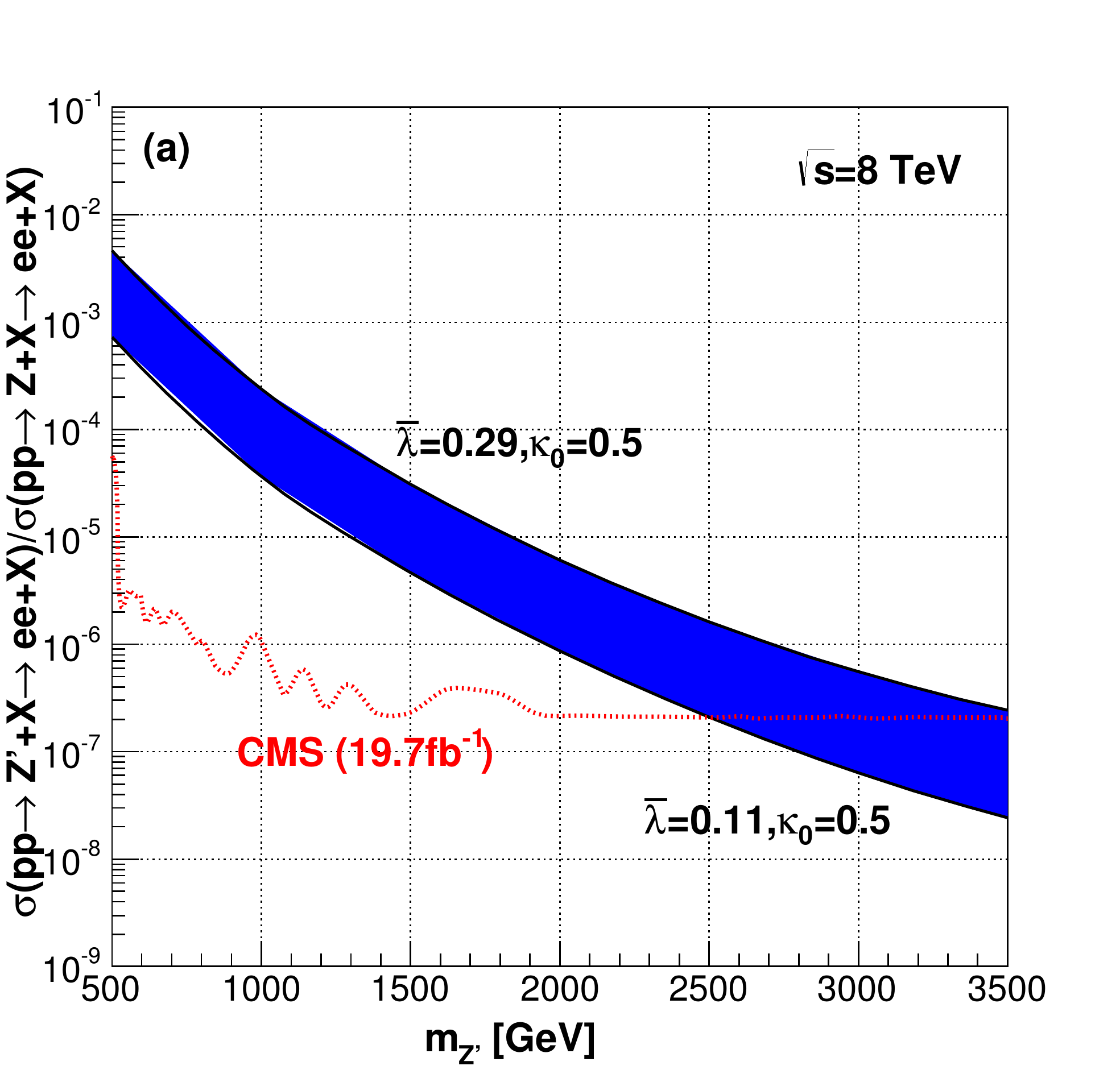}}
  \quad
  \subfigure{\label{fig:LHC:13}\includegraphics[width=0.45\textwidth]{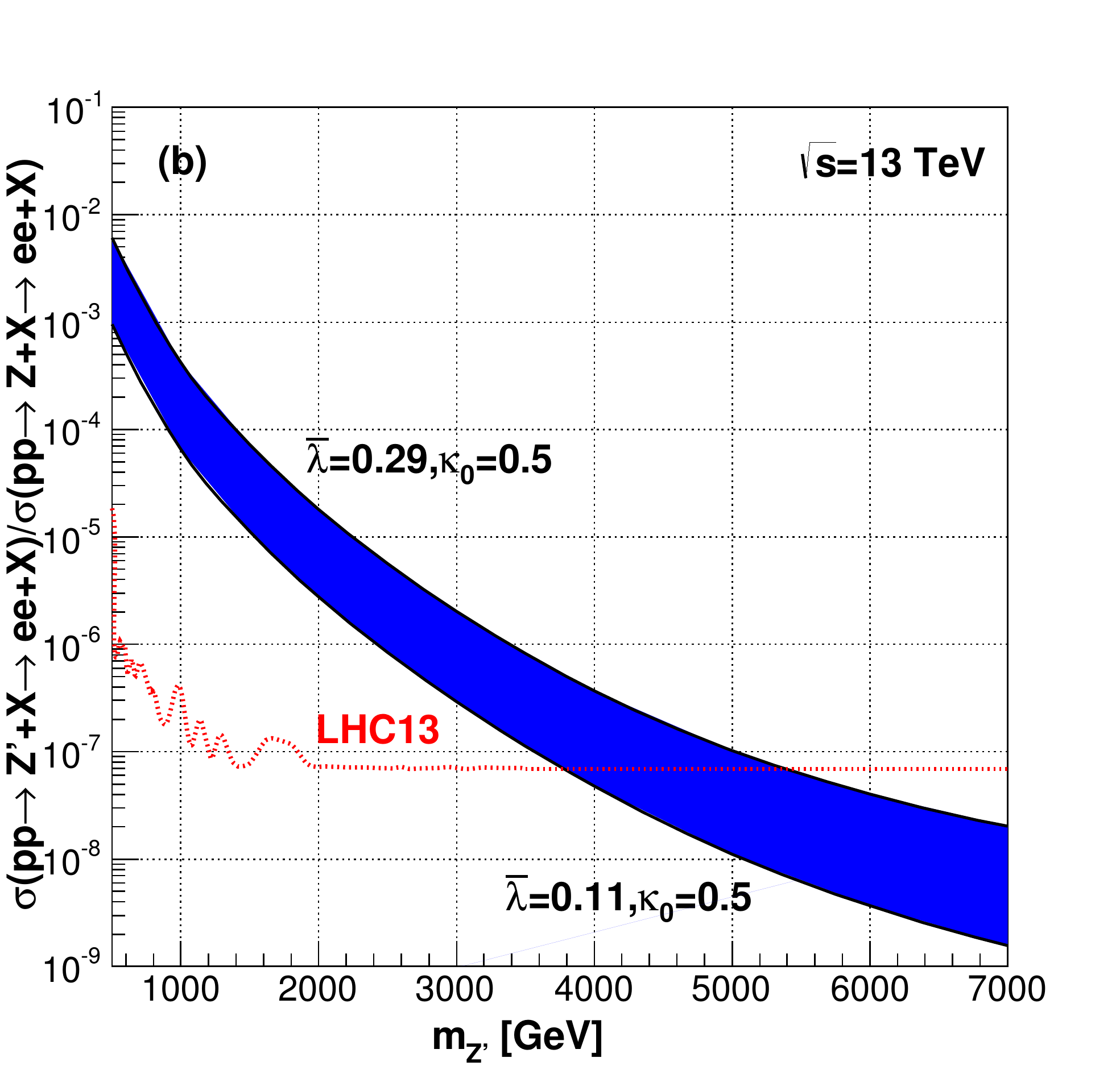}}
  \caption{\baselineskip 3.0ex
    The cross section ratios of $pp \to e^+e^-$ through $\Zp$ and $Z$ production for the two benchmark points of $\bar{\lambda}$ and $\kappa_0$, i) $(\bar\lambda,\kappa_0)=(0.11,0.5)$ and  ii) $(\bar\lambda,\kappa_0)=(0.29,0.5)$ at (a) $\sqrt{s}=8\TeV$  and (b) $\sqrt{s}=13\TeV$. The experimental upper limit are shown in red dotted line, which are obtained from (a) the CMS dielectron channel with $19.7\fb^{-1}$ data at $8\TeV$ and (b) the projected LHC sensitivity at $\sqrt{s}=13\TeV$.}
  \label{fig:LHC}
\end{figure}

In Fig.~\ref{fig:LHC}, we show the cross section ratio of $pp\to e^+e^-$ mediated by $\Zp$ and $Z$ respectively with the center-of-mass energy $\sqrt{s}=8$ and $13\TeV$. The cross sections are computed by using \texttt{MadGraph5\char`_aMC@NLO}~\cite{Alwall:2014hca} complemented with the Lagrangian Eq.~\eqref{eq:Lagrangian:Zp} of the MFV $\Zp$ boson. The following kinematical cuts are imposed according to the CMS experiment~\cite{Khachatryan:2014fba},
\begin{align}
  p_T>10 \GeV,\quad |\eta|<2.5,\quad E_T > 33 \GeV, \quad {\rm and}  ~\Delta R>0.3,
\end{align}
which correspond to the transverse momentum, the rapidity, the transverse energy and minimal separation of final state charged leptons. According to the allowed region of the $\Zp$ couplings to lepton and quark shown in Fig.~\ref{fig:PS_kap0_lamb}, we choose two benchmark points
\begin{align}
 \mathbf{i)}\,\,\,& \bar{\lambda}=0.11, \quad \kappa_0=0.5,\nonumber\\
  \mathbf{ii)}\,\,\,& \bar{\lambda}=0.29, \quad \kappa_0=0.5,
\end{align}
where the values of $\bar\lambda$ correspond to the upper and lower limits with $\kappa_0=0.5$ obtained from a $1\TeV$ $\Zp$ boson. The blue-shaded region therefore satisfy the previous combined constraint. In the $8\TeV$ plot of Fig.~\ref{fig:LHC:8}, we also show the present CMS exclusion bound collected from the $19.7\fb^{-1}$ data in the dielectron channel. In addition, the expected LHC sensitivity at $\sqrt{s}=13\TeV$ with integrated luminosity ${\cal L}=100\, {\rm fb}^{-1}$ is shown in Fig.~\ref{fig:LHC:13}, which is estimated with the method in Ref.~\cite{Jung:2013zya}.

From Fig.~\ref{fig:LHC}, we can see that $m_\Zp<2.5\TeV^{}$ is disfavored for the point i) $(\bar\lambda,\kappa_0)=(0.11,0.5)$ from the 19.7 ${\rm fb}^{-1}$ CMS data at LHC run I, while for the point ii) $(\bar\lambda,\kappa_0)=(0.29,0.5)$, $\Zp$ mass smaller than $3.5\TeV$ is excluded.
At the $13\TeV$ LHC with $100\fb^{-1}$, sensitivity to the cross section ratio of $pp \to e^+ e^-$ mediated by the $\Zp$ and $Z$ boson is expected to increase roughly by a factor of 2 compared to the current experimental bound at LHC run I.
For the benchmark points i) and ii), the $\Zp$ mass below $3.8\TeV$ and $5.4\TeV$ are respectively within the sensitivity reach of LHC run II.
%Combining the flavor constraints of shaded region and the LHC exclusion limit at $\sqrt{s}=13$ TeV, {\bf we can see that the shaded region below $4 ~{\rm TeV} <m_{Z^\prime}^{}<5.5~{\rm TeV}$ is within the sensitivity reach by LHC run II for $0.11<\bar{\lambda}<0.29$ with $\kappa_0=0.5$.}

\section{Conclusions}\label{sec:conclusion}

In this work, the general family nonuniversal $\Zp$ model with a mass of TeV scale has been investigated adopting the MFV hypothesis to avoid potentially large tree-level FCNCs in quark sector. We also extend the general scenario to lepton sector. Considering the MFV $\Zp$ couplings with fermions, their impacts on various low and high energy processes have been studied in detail. It is found that lepton LFV decays $\mu \to 3e$ and $\mu \to e\gamma$, $\mu$-$e$ conversion in nuclei, $b \to s \ell^+ \ell^-$ transitions, $B$ and $K$ meson mixing, and the LEP processess $e^+ e^- \to f\bar f$ are more sensitive to such a $\Zp$ boson.

After a combined constraint from the current experimental data, the allowed parameter space is derived. We find that the MFV $\Zp$ boson can explain the current anomalies in $b \to s \ell^+ \ell^-$ transitions with a non-vanishing couplings $\kappa_1$ and $\bar\lambda$, which controls the FCNC couplings in quark sector and flavor conserving couplings in lepton sector. The implications of these two nonzero couplings are investigated for various processes. In particular, the mass difference $\Delta m_s$ in $B_s$-$\bar B_s$ mixing and the parameter $\abs{\varepsilon_K}$ in $K^0$-$\bar K^0$ mixing are enhanced by more than $3\%\sim 5$\%, which are smaller than both experimental and theoretical uncertainties in the near future. At the same time, the various LFV $B$ and $K$ meson decays are less enhanced and less promising in the near future. In addition, our predicted lower limit on the branching ratio of $\mu \to 3 e$ is $\mathcal O(10^{-14})$ for $\lambda_1 /m_\Zp >0.01\TeV^{-1}$, which is much higher than the expected experimental sensitivity of near future.

At the LHC, the $\Zp$ boson can mediate clear leptonic signal through Drell-Yan channel $pp \to \Zp \to \ell^+ \ell^-$. After considering constraints from the LHC run I data, there are still allowed parameter space in our scenario. It is noted that, the lepton flavor conserving coupling $\bar\lambda$ acquired a nonzero value to explain the current $b \to s \ell^+ \ell^-$ anomaly. For a given quark flavor conserving coupling $\kappa_0$, the cross section of $pp \to \Zp \to \ell^+ \ell^-$ is lower bounded, which make the collider signatures at LHC very predictive. In the near future, if the $b \to s \ell^+ \ell^-$ anomaly persists, direct searches at the LHC run II may have a good opportunity to distinguish various $\Zp$ scenarios, together with high precision measurements at Belle II and LHCb.

\section*{Acknowledgments}

This work is supported by the NRF grant funded by the Korean government of the MEST (No. 2011-0017430) and (No. 2011-0020333). We thank Javier Virto for providing the data in Ref.~\cite{Descotes-Genon:2015uva} and discussions.

\baselineskip 3.0ex

%\bibliographystyle{JHEP}
%\bibliography{draft}

\end{document}